\documentclass[twocolumn]{article}                    
  \usepackage{graphicx}                % allow us to embed graphics files
  \DeclareGraphicsExtensions{.pdf,.png,.jpg,.jpeg} % for pdflatex we expect .pdf, .png, or .jpg files

\graphicspath{{figures/}{pictures/}{images/}{./}{img/}} % where to search for the images

\usepackage{microtype}                 % use micro-typography (slightly more compact, better to read)
\PassOptionsToPackage{warn}{textcomp}  % to address font issues with \textrightarrow
\usepackage{textcomp}                  % use better special symbols
\usepackage{mathptmx}                  % use matching math font
\usepackage{times}                     % we use Times as the main font
\usepackage{cite}                      % needed to automatically sort the references
\usepackage{tabu}                      % only used for the table example
\usepackage{booktabs}                  % only used for the table example
\usepackage{subcaption}

\title{Steps towards prompt-based creation of virtual worlds}

\author{Jasmine Roberts\footnote{: equal contribution, e-mail: jar072@ucsd.edu}\\ %
        \parbox{1.4in}{\scriptsize \centering UCSD \\ Microsoft} %
\and Andrzej Banburski-Fahey$^\ast$ \\ %
     \scriptsize Microsoft %
\and Jaron Lanier\\ %
     \scriptsize Microsoft}

\begin{document}

\maketitle

%% Abstract section.
\abstract{ 

Large language models trained for code generation can be applied to speaking virtual worlds into existence (creating virtual worlds). In this work we show that prompt-based methods can both accelerate in-VR level editing, as well as can become part of gameplay rather than just part of game development. As an example, we present Codex VR Pong which shows non-deterministic game mechanics using generative processes to not only create static content but also non-trivial interactions between 3D objects. This demonstration naturally leads to an integral discussion on how one would evaluate and benchmark experiences created by generative models - as there are no qualitative or quantitative metrics that apply in these scenarios. We conclude by discussing impending challenges of AI-assisted co-creation in VR. 

}

\section{Introduction}

World mythology often includes shapeshifter figures that transform as part of their interactions with each other and the human world.  Pre-digital themed gaming, such as Dungeons and Dragons, often included more transformative gameplay than has become typical of digital gaming.
Some of the early hopes for social VR when it was first demonstrated included allowing people to spontaneously create transformations, in the hopes of bringing about speculative future modes of communication, which were sometimes called “post-symbolic” \cite{kelly1989interview, lanier92}. In this paper we discuss recent technological advances that address these ideas.

The recent progress in self-attention based AI models has led to surprisingly good text and image generative models. Multimodal text-to-image models, like DALL-E 2 \cite{ramesh2022hierarchical}, Midjourney \cite{holz} or Stable Diffusion \cite{rombach2022high} are raising concerns about displacing concept artists and have already won at least one major art competition \cite{roose_2022}. Large Language Models (LLMs), like GPT-3 \cite{brown2020language}, are not only generating very convincing text completions, but have recently become capable of generating code with models like OpenAI Codex \cite{chen2021evaluating} or AlphaCode \cite{li2022competition}. We propose in this paper that these capabilities can be combined to allow "speaking the world into existence", or taking natural language descriptions and turning them into interactive visual scenes within a game engine. In particular, this has the potential for allowing authoring Virtual Reality (VR) experiences from within the headset, as well as allow completely novel modes of gameplay.

Development of VR content is a slow and challenging process, often requiring a whole dev team for a short experience. This is even more pronounced when we  allow generation of content by the user at runtime. While there have been attempts at allowing in-VR level editing (for example EditorXR \cite{unity-technologies}) they have not seen wide adoption due to their limitations.  Integrating Codex with a game engine however should allow for a much easier real-time natural language interface for interactive scene creation. This should in principle allow us to not only create game levels on-demand, but also allow non-coding users to speak training or educational scenarios with much less effort and time expenditure. We discuss this in more detail in Section \ref{sec:creation}. We then discuss several potential high-impact scenarios that could result from this work in Section  \ref{sec:scenarios}.

To show that prompt-based creation can become part of gameplay rather than just part of game development, we demonstrate the first VR game with non-deterministic game mechanics powered by OpenAI's text generative models, by integrating them with the Unity game engine \cite{technologies}. In reference to one of the first video games ever made  \cite{Harrison1964COMPUTERAIDEDIS}, we built a surreal tennis game, in which the players can transform both the paddles and the ball into any 3d objects. These transformed objects then interact in semantically sensible ways that were not predetermined by the developer, for example a ball transformed into an egg colliding with a frying pan results in a fried egg. We describe the details of this demonstration in Section \ref{sec:vrpong}. Along the way of building this demonstration, we experimented with building a rudimentary Holodeck for global environment generation, as well as more fine-grained modification of user's hands into tools.  In attempting to speak virtual worlds into existence by applying Codex in a game engine, one runs into a host of technical challenges. This spans the breadth from latency issues, providing feedback to the generative model, automating animations, to 3D model repository choices. We consider some of these topics in Section \ref{sec:technical}.

A natural question that arises in this line of work is how we should objectively evaluate the quality of completions that we receive from these AI models in this open-ended setting. A connected question is whether it is possible to have an automated scoring system in the setting presented here, in which both the game objects as well as the rules of any game are in principle mutable. While some quantitative measures exist for specific problems in text generation, we lack any clear objective benchmarks in the domains this paper focuses on. We discuss some potential ways forward in Section \ref{sec:evaluation}.

% Traditional 3D game design is significantly more challenging than 2D, and this is even more pronounced when attempting to generate content at runtime. While recent models like DALL-E 2 \cite{ramesh2022hierarchical}, Midjourney \cite{holz} or Stable Diffusion \cite{rombach2022high} could conceivably be used to generate 2D textures that could be animated with other tools \cite{wu2022nuwa}, 3D content generation from scratch is very nascent . This leads us to integrating Codex with the Unity game engine \cite{technologies}, which has the potential for unbounded creativity from the user/player, but introduces nondeterminism that carries the risk of removing all sense of agency from game design. In this work, we choose to study a restricted semi-deterministic setting with some implemented hard-coded mechanics while trying to not compromise our investigation of LLMs in-game. In homage to the first video game made \cite{Harrison1964COMPUTERAIDEDIS}, we developed a surreal game of VR tennis for two, in which players can both change the ball as well as their paddles to arbitrary 3D objects and allow for semantically relevant interactions between these morphed objects. 

\section{Related Work}

Previous immersive authoring research has proven benefits to runtime authoring  feature sets within tools. LevelEd VR \cite{level_ed} created a generic level designer to support the block out process for game developers. A key benefit of their system was the developer’s ability to quickly edit and test the level with live gameplay. They also indicated that such a system that does not require imperative programming, 3D modeling, or other expertise can be released to end users for the creation of user generated content once the game is released. 

Flowmatic \cite{flowmatic} introduced the ability to author interactive and discrete events for which outcomes could be specified. Some examples are collision events, what happens when the object is instantiated, destroyed, or directly interacted with by a user.  The ability to abstract, create, and reuse such behaviors in an immersive authoring environment allowed the creation of more complex interactions without having to switch between developing and testing. Users could also easily understand which objects of the scene their changes were impacting.  One overlooked benefit of quickly editing an object and switching to testing is that the author has a better understanding of the datum and orientation of the object and can make rapid fixes to what are normally complex interactions when dealing with quaternions and the proper scaling and placement of an object. One drawback of Flowmatic is that it is still designed with programmers in mind. Further abstraction would be needed for a standard user to quickly utilize such a tool if the end result was for use of user generated content. With Codex, we have added a necessary  layer of abstraction to empower not only programmers but also players with the ability to quickly generate dynamic gameplay results.

 Initial investigations have shown that text-to-code models can write short programs and match the performance of average competition coders \cite{li2022competition}, as well as solve university-level math problems \cite{drori2022neural}. In this work, we demonstrate the potential for real-time co-creation applications in VR of such generative models. One of the first co-creation applications of Codex is GitHub Copilot, in which the AI model co-creates with a developer by completing code from comments \cite{github}. Previous work has shown that it is possible to create NPCs capable of following natural language orders \cite{volum2022craft} or text adventure games \cite{walton}. The first demonstration of using Codex to create 3D scenes was that of Codex-Babylon \cite{Codex-Babylon}, in which Codex is used to create and operate on 3D models with the Babylon.js web-based rendering engine \cite{babylonjs}. This was then applied by FrameVR to make a demonstration of speaking a virtual lesson about the moon into existence \cite{framevr_build}.

Of high importance to the endeavor to speaking worlds into existence is the ability of loading in arbitrary 3D models into the scene. While in our demonstration we use API calls to Sketchfab, having end-to-end generative models would provide unlimited levels of customization. The two main technologies that could achieve this are Neural Radiance Fields (NeRF) \cite{Mildenhall20eccv_nerf, Garbin21arxiv_FastNeRF}, and direct text-to-mesh generation \cite{khalid2022text}. While until recently NeRFs required photos of the reference object from multiple points of view  to generate a mesh, recent work at Google has resulted in Dreamfusion \cite{poole2022dreamfusion}, which generates NeRF models from a single 64x64 image generated by a diffusion model. Another recent approach in generating meshes is GET3D \cite{gao2022get3d}. While these are very promising steps, it is so far unclear how one would automatically animate arbitrary meshes generated by  these models -- there is  however some progress in animating NeRF humanoid avatars \cite{2021narf, peng2021animatable, chen2021animatable}.

\section{Prompt-based creation in VR}
\label{sec:creation}

Virtual Reality and related modalities would benefit from prompt-based content generation, analogous to what Dall-E has done for images or GPT for text.  Indeed, it might benefit more than other domains, for two reasons.

The first is that content generation has proved to be a critical bottleneck for VR.  VR is a complex, heterogeneous medium that resists standardization of development practices.  For instance, excellent educational content (such as experiential demonstrations of relativity \cite{doi:10.1119/10.0001803}) has been developed in VR, but despite effectiveness, these resources tend to not be widely used.  There are multiple reasons, but a persistent reason is that it is hard to keep content updated to be compatible with changes in hardware (especially user interaction hardware), real time architectures and channels, and other elements.

A prompt-based method that reduces the cost of generating VR content might not only spur the appearance of new content by reducing the time and cost of development, but might also make it more feasible to make already-existing content continue to be available.

A second motivation to explore prompt-based generation in VR is that spontaneous user-generated content in the course of a VR experience has been imagined as a core element of VR from its inception \cite{kelly1989interview, lanier92}.  There are several reasons for this.  One is that VR has been imagined as a form of “social lucid dreaming” in which people could make up a shared environment as a form of interaction.  This was sometimes referred to, starting in the 1980s, as “post-symbolic communication”, meaning that a people would make up the world as experienced by others instead of describing it symbolically with words.

The spontaneous generation of worlds with words has been a part of human culture since the dawn of recorded history.  Mythologies all over the world include shapeshifting characters that transform themselves. The "Wizards Duel" from the 1963 film The Sword in the Stone \cite{disney} perfectly depicts a battle of wits involving such transformations. Fantasy gaming before computers were widely available, such as in Dungeons and Dragons often included frequent world invention in the course of gameplay. Fictional depictions of VR, such as in Star Trek’s Holodeck, or The Matrix movies, tend to depict characters asking for the content and dynamics of the world to change in response to prompts.  Digital fantasy gaming has been the only type of fantasy gaming in which this level of creativity is not typical.
Furthermore, practical uses of VR would often be made economical by a prompting methodology.  For instance, a physical therapist might prompt for a virtual exercise machine or environment that is tuned the needs of a patient with an unusual case.  Emergency first responders might be able to train in a simulation of a specific, unusual incident while on the way to a site.  These types of applications of VR are not possible now because of the time and expense of development.

The idea of creating virtual environments from within the VR headset has a long history. In the 1980s, Jaron Lanier proposed that virtual musical instruments might someday play the world into existence \cite{lammers_1986}, or that people might dance the world into existence.  These are still plausible interaction ideas and should be pursued.

At present, however, textual prompt-based methods have achieved at least initial utility in text, code, and image generation.  Therefore it is natural to investigate spoken prompts as a method to “speak the world into existence”, or change the contents and dynamics of a virtual world while one is experiencing it.

There are several recent advances that point to this possibility, though it is important to distinguish the requirements of VR from gaming and other similar or overlapping modalities.
These overlap requirements for VR, but VR requires additional capabilities, including:
\begin{itemize}
    \item Objects in VR must be rigged, meaning that internal joints and other degrees of freedom must be present and actionable in the environment.  A good example is a library of rigged avatars \cite{10.3389/frvir.2020.561558}.  Mesh generation does not automatically produce rigged objects.  Indeed almost all mesh content that is readily available by any method is not rigged.
    
    \item Objects in VR must have interaction zones or other interaction contingencies.  For instance, an avatar's hand should be able to pick up certain objects, but typically the avatar's feet should not.  Furthermore, the picking up of a thing might become a significant dynamic, pseudo-physical model in its own right.  Virtual objects often have elements such as handles, buttons, or other subsections designated for interactions.
    
    \item There must be a degree of semantic coherence to how objects can interact.  One should be able to pick up a virtual suitcase, but not the water in a pond, or at least not in the same way.  At the same time, interactions must not be too rigid.  It should be possible for surprising interactions to emerge between objects or other elements.
    
    \item 	Coherent whole scenes should be prompt-able.   A user should be able to ask for a forest, or even a tropical Halloween forest, and an ensemble of related elements, like trees, animals, creeks, and boulders should assemble and interact with one another appropriately.
    
    \item Users should be able to refer to what has been created in order to modify it, using both words, descriptions, pointing, and other modalities.

\end{itemize}

This is an ambitious list of requirements that will not be met soon, but we have defined a course or research to touch on them and show incremental progress.
\begin{figure*}[ht!]
  \centering
  \begin{subfigure}{0.49\linewidth}
  \centering
  \includegraphics[width=.98\linewidth]{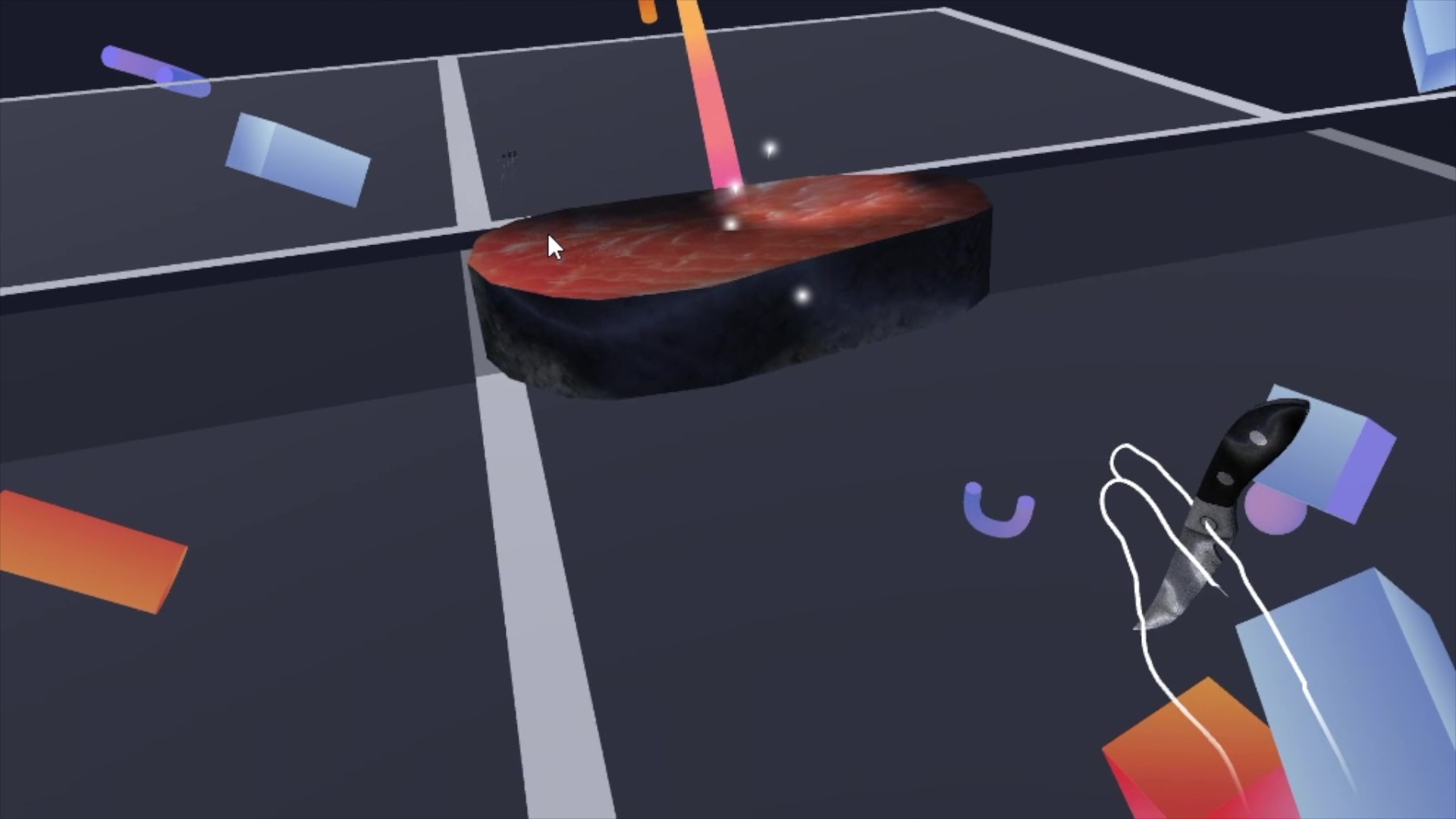}
    %\caption{Ball $\rightarrow$ salmon, paddle $\rightarrow$ knife}
\end{subfigure}
\begin{subfigure}{0.49\linewidth}
  \centering
  \includegraphics[width=.98\linewidth]{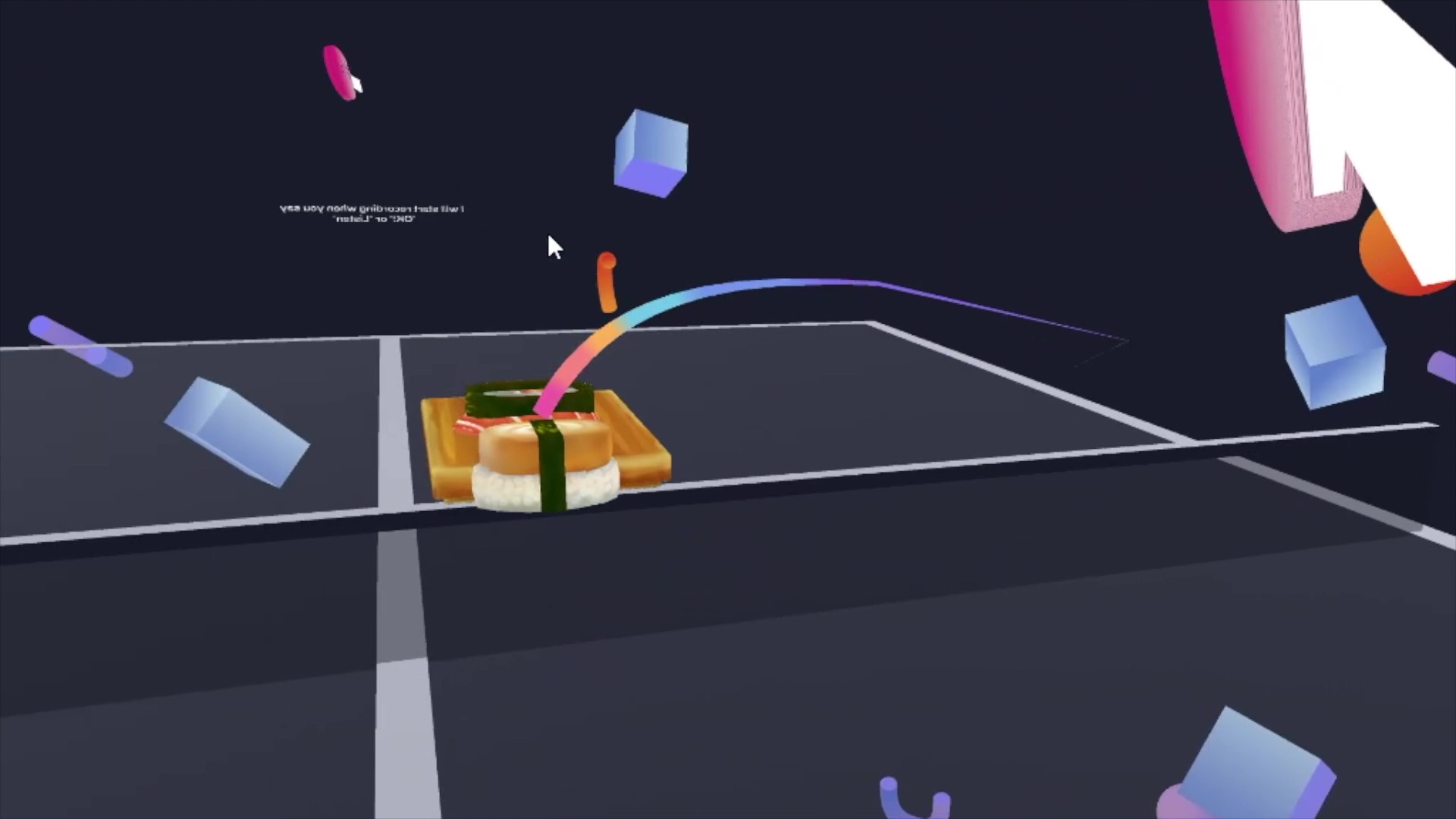}
    %\caption{Collision results in sushi}
\end{subfigure}
  \caption{Codex VR Pong as an example of prompt-based gameplay mechanics: the ball and the paddle can be transformed arbitrarily and can interact in unscripted ways. Here, a ball was transformed into a salmon steak, the paddle into a knife, and upon their collision the salmon turns into sushi.}
  \label{fig:teaser}
\end{figure*}
\subsection{Potential scenarios}
\label{sec:scenarios}
The idea of using the progress in natural language processing (NLP) for spontaneous content creation in virtual reality and metaverse sounds very appealing, but what kind of specific applications can we expect? While it is clear that giving users/players a sandbox in which they can speak anything into existence is crucial to the eventual success of the metaverse project, we want to discuss some more concrete applications.

\begin{itemize}

    \item A fantasy roleplaying Game Master could describe the environment to their adventuring party and have it appear in real-time, gaining more detail whenever the players ask specific questions about parts of the dungeon they find themselves in. They could also speak traps into existence, or foes for the players to face.

    \item A physics teacher could prepare an immersive lesson in relativity and astrophysics in VR the evening before giving the class, specifying the rules under which objects in their demonstration behave. Such a demonstration would currently be unfeasible to implement by a single person in a short time frame. 

    \item In a virtual environment, students within a lab setting could inspect and work with materials or machines which would otherwise be outside the school's budget or unsafe to handle in the physical classroom. They could then utilize voice requests to alter the experiment and evaluate the outcome. 

    \item In a virtual simulation of a real life emergency situation, the training instructor for first responders could dynamically introduce otherwise infeasible hazards to assess the trainees ability to adapt to unexpected situations. For example, in firefighting training an instructor could cause parts of a stairway to collapse or spawn artificial victims in need of rescuing that could be made to respond in a panicked and unsafe ways.   %The trainee could then instantiate objects needed to circumvent and handle the hazard.  
    
     \item A warehouse designer could utilize such an application to request common warehouse materials such as shelves, packages, bins, conveyors, fork lifts, loading gates, and personnel. They could then place the objects to evaluate throughput, flow, accessibility, and safety of the design. 
    
    \item Consultants in an ideation session could use voice prompts to generate multiple interactive 3D assets with the intent to inspire more prompts with a novel output. The process would be repeated until there was convergence of agreement for a critical path and further discovery.

\end {itemize}

Speaking content into existence can also enable novel gaming mechanics. In our implementation, Codex Pong, a surreal multiplayer virtual reality game of pong, we control  for user input but allow output randomness. The output randomness is offloaded to GPT-3. Output randomness creates unknown outcomes, which results in more compelling narrative moments. The details of the implementation of this example can be found in the following section. We expect that this demonstration should open up the doors for more open-ended transformative experiences in gaming, ones for which the developers do not have to prepare for beforehand, opening the potential for truly surprising interactions.

\section{Demonstration of prompt-based creation}
\label{sec:vrpong}
Here we discuss how we implemented Codex VR Pong, as well as a couple of other noteworthy experiments along the way.

\subsection{Codex VR Pong}

\begin{figure}[b]
  \includegraphics [width=\linewidth]{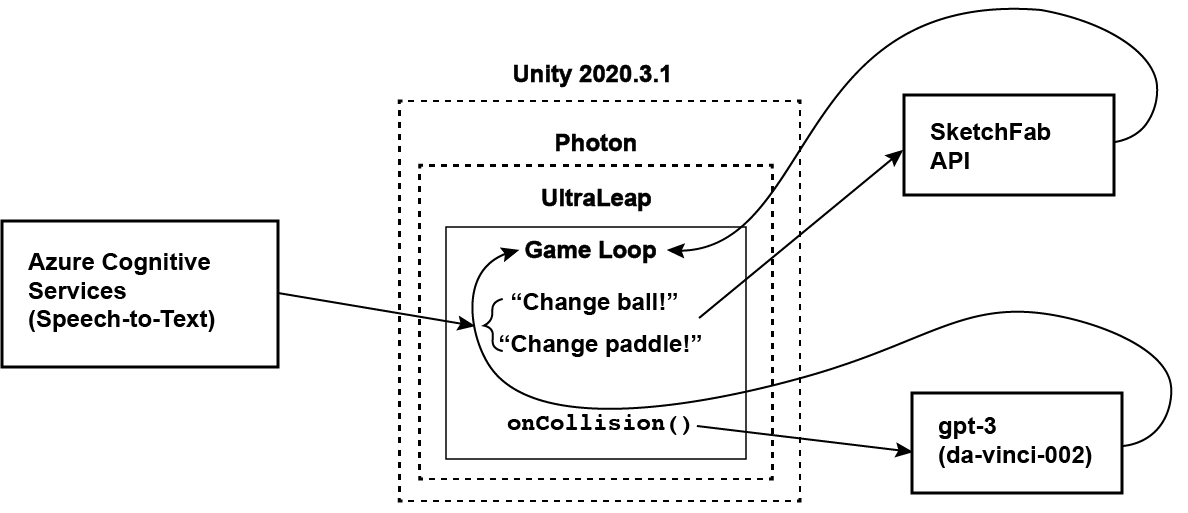}
    \caption{Codex VR Pong system diagram}
    \label{fig:diagram}
\end{figure}

To create a multimodal 2-player game of surreal tennis we need a myriad of technologies working (see Figure \ref{fig:diagram}). The first of these is the integration of Codex with Unity. While Codex is most capable in Python and Javascript, it has been fine-tuned on majority of GitHub, which includes $C\#$ code used in Unity projects, as can be evidenced by simple experiments with the OpenAI API. The immediate obstacle one encounters is that $C\#$ code needs to be compiled, and by default Unity does not ship with a compiler that would allow evaluation of code generated at runtime. Fortunately, the open source Roslyn $C\#$ compiler has been successfully implemented in Unity previously \cite{interactive}. To integrate the OpenAI API, we expanded a previous attempt to build a .NET wrapper \cite{welsh}.

We used both text generation through GPT-3 and code generation through Codex to achieve the experience. To enable creative interactions, we used the {\it text-davinci-002} model with temperature $T=0.5$, and provided it with a contextual prompt:  "This is a magical game like ping pong, in which the players can change both the ball and their paddles, when the transformed ball object hits the transformed paddle, it changes the ball according to how you'd expect those two objects to interact." Working in the few-shot regime, we provided several examples and then prompted the actual interaction within the game, e.g. "When water collides with fire, it spawns". From this sentence, GPT-3 predicts the most likely following word (in this example, it returns "steam"). See the table for more examples and the Appendix for the full context provided to GPT-3.

The output from GPT-3 is then be converted to a 3D model and imported in the scene as the replacement of the previous ball. Every output is saved to a text file for future reference.

We then separately used  {\it code-davinci-002}  with $T=0$ to obtain scripts implementing model downloading, replacements and rescalings (possible due to Codex's real world knowledge, which includes the usual size of common objects), and any other requests the user might have. While Web extraction tools have perfected algorithms for harvesting 2D media like images and video, for our game we needed a database with a sizable number of 3D assets. We dynamically loaded models at runtime via the Sketchfab API (which contains over 3 million models). All the models that contain the requested object get ordered by their number of likes and then a random object with the lowest vertex count gets selected as the model to download. This way, download speed and model quality are being optimized. If one model takes more than 5 seconds to load, the next model in the list will start downloading to avoid failed downloads and extended loading times. Once the download is finished and the object is successfully spawned, a remote procedure call is sent to inform the other player's game instance of the changed 3D model. This way, the other client can download the same model with the download link to synchronize the models. To prevent objects from being too large or small due to the different proportions of the downloaded models, we chose to calculate the bounding boxes of the objects and re-scaled each object accordingly.

For user input we used voice commands via Azure Cognitive Services Speech to Text \cite{azure}. The system continually listens for phrases to initialize user input: “Change ball” changes the model of the ball/projectile, “Change Paddle” swaps out a paddle model affixed to the user’s hand with an object the user specifies, with the resulting code completion displayed on a panel overhead the user.  We use the UltraLeap Stereo IR 170 for tracking players’ hands and Photon is for multiplayer networking. 
. 
\begin{table}[tb]
\scriptsize%
	\centering%
	\begin{tabu}{%
	ccc
	}
  \toprule
   Ball Object & Paddle Object  &  \textbf {Output}   \\
  \midrule
	 salmon & knife &  \textbf{sushi} \\ 
 fried egg & time &  \textbf{rotten egg} \\ 
 fire & ice &  \textbf {water} \\ 
 family & time &  \textbf {memory} \\ 
 memory & disaster &  \textbf {PTSD} \\ 
 pineapple & banana &  \textbf {smoothie} \\ 
 apple & tennis racket &  \textbf {apple pie} \\
 dinner & trash can & \textbf{maggot} \\
  \bottomrule
  \end{tabu}%
\caption{Example interactions between the morphed ball and paddle objects in Codex Pong.}
\end{table}

% “Pull” changes the inertia of the ball towards a player’s paddle

\subsection{Other experiments}

In building Codex Pong, we experimented separately on how much we could push Codex in global scene generation and more fine-grained interactions  with the player's hands. 

\paragraph{Holodeck}

\begin{figure}[h]
  \includegraphics [width=\linewidth]{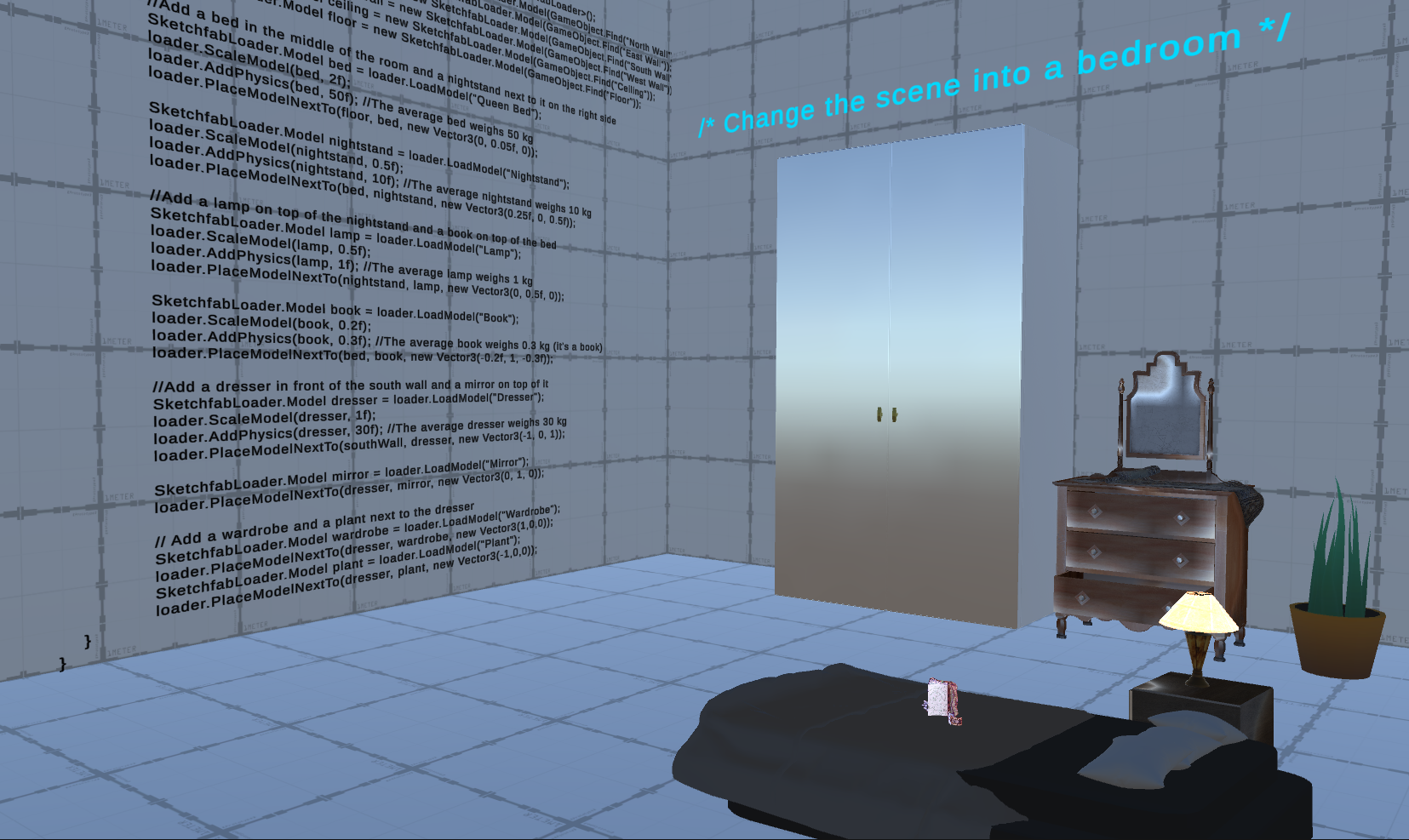}
    \caption{Holodeck experiment in which we spoke a bedroom into existence by the simple command "Change the scene into a bedroom". Codex is capable of generating complex scenes, but has difficulties in placing the objects in natural orientations or positions without further guidance.}
    \label{fig:holodeck}
\end{figure}

One of the long imagined applications of VR is the Holodeck from Star Trek, which is a mixed reality room capable of generating convincing simulations of arbitrary environments and virtual characters given a user's voice commands. We implemented a very simple version of Holodeck as a test of how complex environment generation Codex is capable out of the box -- that is without any special fine-tuning.  We provided codex with a simple description of the goals of the Holodeck experience (see the Appendix) and provided it an example of loading two objects into the scene. We then asked it to change the scene into a series of different environments. We display the generated code on one of the surfaces of the room. See Figure \ref{fig:holodeck} for an example of a generated bedroom. 

We have found interestingly that Codex can place multiple objects into the scene and arrange them without overlapping, to a highly varying degree of natural placements. Because the objects in Sketchfab are not always oriented consistently, we often receive objects with random rotations. When prompting for very complex scenes, like a jungle, we found that sometimes Codex outputs a reasonable short code, while other times it gets stuck in a sequence of adding similar objects to the scene over and over (e.g. adding over 20 trees sequentially, rather than using a loops to achieve the same quicker). Setting the \emph{frequence penalty} hyperparameter of Codex to a small nonzero value penalizes this repetition and helps to avoid it most of the time. 

Our Holodeck experiment suggest that Codex can indeed generate complex scenes, but does so better with at least a few examples (in the few-shot regime). While Codex can generate a scene with multiple objects with a very short description, we found that we can get richer scenes if we first pass the user's prompt to an instance of GPT-3 with a higher temperature (to achieve more creative answers) and ask it to elaborate on the prompt and make it a detailed step-by-step instruction. Future work should evaluate how much this can be improved by fine-tuning the model specifically for the task of environment generation. Currently, if we provide too many examples in the context, it is very easy to hit the limit of tokens the model can pay attention to, but this should be alleviated with future larger models.

\paragraph{Speaking tools onto hands}
As a predecessor to the Codex VR Pong, we needed to verify the technical capabilities of Codex interacting with the Leap Motion hand tracking and thus experimented with proprioception and embodiment.  We looked at another aspect of AI-assisted co-creation: speaking tools into existence. Codex is capable of understanding natural language commands to spawn virtual objects on all the joints in the hand. We envision that one could speak an array of tools, from a simple paint brush, to a mesh sculpting tool, a mesh cutting one etc. A tool of note we experimented with is a Laplacian deformation of meshes tied to the different joints of the hand -- Codex can be prompted to tie the deformation of any mesh in the scene to  the relative position of two fingertips for example. 

This proof of concept allowed us to overcome the initial technical challenge of replacing virtual hands as the default representation of the user's hand position. In our implementation of Codex VR Pong we utilized this feature  to replace the user's hand with a paddle and then other spawned objects. This functionality could be used in further research regarding embodiment and proprioception. 

\section{Technical challenges \& considerations}
\label{sec:technical}

While implementing our demonstration, we encountered several technical challenges in need of further research. Unresolved problems  diminish sense of agency of the player, or the experience of controlling one’s own actions and their corresponding effects in the environment. In our multimodal example, the clear primary drawback is latency. This comes from two sources: loading the 3D model and the latency inherent in using LLMs. Some other considerations we discuss here include the amount of feedback the generative model receives from its completions, how to deal with animations and the challenges of using compiled languages.

\subsection{Latency}

The overall model loading latency can be impacted by multiple technical factors such as network speed, system resource utilization, and complexity of the instantiated 3D model. While it is apparent that low network speeds and large file sizes increase latency, the trade-offs between user comfort and request completion times are not so obvious. Downloading the model requires significant CPU resources which are already strained by the high system requirements needed to render virtual reality. Once the model is downloaded, the CPU resources are strained again when the model is loaded into memory during instantiation. These necessary processes can reduce the frame rate of the experience significantly below the threshold needed to avoid simulator sickness. 

One way to curtail significant reduction in frame rates is to limit the CPU resources given to the asynchronous threads downloading the model. This increases the total latency between the initial voice request from the user and instantiation of the requested object into the scene. When using voice commands, increased delay times cause the player to feel as though they are not interacting with the system properly. This confusion can lead to the player repeating the input, altering the input, or attempting to cancel the request. Because there are multiple endpoints, parallel requests can create conflicting race conditions and impede service requests. 

Another concern is GPU resources needed to render all objects. Each new model has the risk of causing frame rate degradation. Preventative algorithms could be implemented to assess an objects’ impact to performance prior to downloading and even inform the user that other objects must be removed from the scene before the request can be completed. 

This safeguard could increase total latency of the voice request if an optimal object must be searched for in Sketchfab. These concerns apply to each client utilizing Codex end-to-end. In the networked multiplayer scenario, all these processes must propagate in tandem and be completed by multiple users, who each have varying network speeds. Ideally, the object would wait to instantiate itself until all users have successfully downloaded it. Dips in performance caused by all the aforementioned examples can desynchronize the position of objects for networked players which causes erratic and confusing gameplay results. Another problem introduced by the multi-user scenario is when both users make requests at the same time. Ideally, a queuing system would mitigate if not entirely prevent the situation of doubling the tax on resources introduced by simultaneous Codex requests. 

The other source of latency is that of Codex -- while short completions of a single model download can be fast, requiring Codex to write a whole class implementing complex behavior can often take upwards of 10 seconds. While there is currently no solution to this latency, one possible solution would be to take inspiration from the framework of two systems of reasoning \cite{kahneman2011thinking}, one fast and one slow. One could imagine using a smaller faster model to provide a first draft of the completion, and display the results of that quickly, while allowing the larger language model to continue on a more detailed answer. A big challenge here would be to impose consistency between the two completions.

One potential approach to dealing with the latency is an implementation of a semi-turn based approach seen in some video games, like  Final Fantasy VII Remake for example \cite{mixandjam}, in which a player can slow down the time to allow themselves more time to make tactical decisions. Here, we can temporarily multiply the \texttt{Time.timeScale} in Unity by a small number (e.g. 0.01) whenever the player starts object transformations. This could be accompanied by a visual cue, for example muting the colors. This would both give the player more time to make their decisions, as well as make the process of loading the model less disrupting. Once the model is properly loaded in, the game can speed up again.

\subsection{Compiled vs interpreted languages}

Our implementation of Codex in Unity crucially depends on runtime compilation provided by Roslyn C$\#$. One limitation of this approach stems from the simple fact that in compiled languages like C$\#$, one has to be clear about the code dependencies on external DLL assemblies. Using Roslyn one can preload desired assemblies, but it is impossible to add new ones at runtime. While this does not impact code using only Unity assemblies, it does require the developer to in advance know what APIs the user might want to call -- calling an unloaded API results in the code failing to compile. While for simple interactions this should not pose a problem, it does put a limit on the kind of completions we can make.  

The only clear solution to this problem is to bypass it by switching to an interpreted language like Python or JavaScript, rather than a compiled one. There are currently no working solutions for this in Unity, but there exists an extension for Unreal that uses JavaScript \cite{ncsoft}. An alternative to this would be switching to a web-first solution like Babylon.js \cite{babylonjs}, which is a web-based open source rendering engine powering experiences such as FrameVR \cite{framevr}.

\subsection{Feedback}

One of the big limitations in applying Codex to the type of content creation we consider here is the lack of feedback the generative model receives about its completions. Currently, a user's prompt leads to a completion, which may or may not implement what they had in mind. While it is expected a user might have to iterate with the generative model several times till they achieve a satisfactory result, this is complicated by the disconnect between the generated code and the resulting visual scene. In our implementation Codex calls the Sketchfab API to download models, but it does not "see" the resulting scene -- this sometimes results in unexpected behavior, often stemming from the fact that the downloaded model had one of their tags matching the search criteria, despite representing something very different. For example, a user might want to place a table in the scene, but the downloaded model is that of a whole house, that just so happens to also include a table. 

We see two potential solutions to this problem, which could be used in tandem: closing a feedback loop, or using improved 3D model datasets. 

In the first one, note that in situations where the code describes the full visual result, Codex is capable of reasoning about the result and improving on it, for example through its edit functionality \cite{bavarian_2022}. Whenever Codex calls an external API for a visual effect and does not receive a full description of the results, there will be disconnect. This can potentially be alleviated if we combine Codex together with a multimodal text-vision model, for example CLIP \cite{radford2021learning} or Florence \cite{yuan2021florence}. Once Codex completion is compiled, one could capture the result as a screenshot and feed that to the vision model, asking it for a textual description of the scene. This then could be fed back to Codex for potential improvements without the need for the user to intervene. This could be iterated several times until the completion is deemed "close enough" to the original prompt, or until the user is satisfied with the result and interrupts this loop. Note that this approach could introduce quite a significant additional source of latency, and might be more suitable in less time-sensitive applications.

Another option of improving feedback to Codex would be to curate a 3D model repository with very carefully crafted metadata. Currently, the models we load in from Sketchfab are community created and do not have standards for this -- they come with user-decided tags that sometimes do not accurately  match the model. A more useful repository of models would have a short description of the object that could be passed to Codex prior to compiling the results and iterated on before any download would happen. Additionally, such models should be provided with more useful metadata, such as the description of the extents of a bounding box, whether the shape is made up of one or many segments, how many connected components there are, description of any included animations, and should provide the default rotation in natural language. A curated dataset like this could also be useful in keeping a reasonable limit on the actual size of the model to improve download speeds. It is interesting to note that much of this could be automatized if the models came from an approach like Dreamfusion \cite{poole2022dreamfusion}, in which a diffusion model powers a NeRF in generating a 3D model.

\subsection{Animations}

So far we have mostly discussed loading in static 3D models and at most making them dynamic through transform updates. For truly immersive VR experiences, we need to also allow for physically realistic interactions and animations. The former can be automated rather straightforwardly by including appropriate colliders and rigidbody properties in Unity, or adding more complicated physics API calls, both of which Codex is capable of already and which we included in our experiments. Automating animations however is a more challenging topic, worth significant future research. Here we discuss the pipeline for achieving this.

While Codex can easily load in animations that have been appended to the 3D model (as is in the case with many Sketchfab models), creating a scene with an animated model transitioning between different animations in response to various events and filling out an animator controller at runtime with Codex requires very careful prompting by the user currently. What about the case where a model comes without premade animations?

For a fully solid mesh, we would need to automate three steps: segmentation, rigging and the animation itself. Segmenting an arbitrary mesh and then rigging it would require a vision and code generating multimodal model, which is an active area of research. For humanoid models however, there are existing methods for automated rigging \cite{actorcore}, and work in \cite{peng2021animatable, chen2021animatable} has shown that it is possible to generate rigged NeRF models from scratch. For such generated models, as well as avatars from libraries like Microsoft Rocketbox \cite{10.3389/frvir.2020.561558}, Codex can write code that assigns animations from a common pool to them (whether from Rocketbox, or from Mixamo \cite{adobe}).

\subsection{Other considerations}
It is also important to consider that when developing games it may be necessary for more elaborate implementations than Codex provides using improved API calls, wrapping logic into classes/functions, and better kinematic algorithms. This is where human game developers still surpass AI models like Codex. Without additional fine-tuning, Codex often produces sub-optimal game mechanics in many cases. While it is possible to steer Codex by providing it more examples, this is limited by the size of the attention context window and it is easy to run out of tokens in generating more complex code (this should be alleviated however by larger models in future). 

Currently VR experiences are reliant on maintenance of external APIs and SDKs. In the event of deprecation, the experiences become inoperable. In our example, the Sketchfab API can easily be replaced with one of the generative 3D models cited in related works. Furthermore, the integration of Codex into the game engine means that the user should be capable of changing the API used at runtime.

\section{How to evaluate the quality of completions?}
\label{sec:evaluation}

% (Note: factoring time and cost is addressing the viability of co-creating gpt-3, but is it addressing the quality of completions and the generated results?)

A serious problem in prompt-based content generation of any kind is criteria for success.  In the case of text models, like GPT-3, there tends to be a cross between the old idea of the Turing Test and commercial viability.  If people accept a system, it is deemed successful.  In many cases, this implies that users are not aware that the tool was used at all, that there was not a real human.  For image-oriented systems, like DALL-E, there is a somewhat different criterion, in which users enjoy being surprised.  This suggests that users are aware of the presence of the tool. Codex and other code generation systems are judged on whether they reduce the costs and time of coding, also with the awareness of a human programmer.

There are approximate percentages of success applied to the outputs of these models.  For instance, it might be said that a code generating program produces useful results 60\% of the time , but this cannot be in general a robust measurement since there is a subjective component and the circumstances are endlessly varied and never completely described. For example, Codex achieves 70.2\% accuracy on HumanEval \cite{chen2021evaluating}, which is a dataset of programming tasks, in which success is measured against test cases. However, it is much less clear how to evaluate Codex completions on open-ended tasks, such as the ones discussed in this paper.

In evaluating the quality of the completions and the virtual worlds generated by these models, we have to face the fact that there is no objective way of evaluating virtual creations beyond a meta-review of a collection of user scores and reviews. Even in evaluations of creativity of completions, we can only speak of how surprising and efficient they are for a given audience, as measures of creativity are highly subjective \cite{buildings11010029}. This also means that there is no clear objective way of scoring any games using the kind of novel game mechanics that LLMs make available. The only clear automated approach for scoring we see is to use another instance of a LLM that evaluates how interesting or creative a given play sessions was.

In a research environment, it is always desirable to make criteria as rigorous as possible so that there is hope of repeatability.  It can be argued, however, that the very large scale of deployment of current prompt-based systems produces a large enough $N$ that informal criteria like the ones above become sufficient.

How should we evaluate the success of prompt-based VR?
We proposed two motivations for prompt-based VR: lowering dev time and cost, and bringing spontaneous creation into VR experience.

If the motivation is dev time and cost, there are two clear directions for criteria.  One is that we could compare the development of a given virtual world or experience when produced with or without prompts (meaning using any other dev tools).  In this case, the methodology of studying prompt-based VR resembles that for Codex more than for DALL-E. This would suggest running a "low end" study in which a group of non-experts (not developers) would use the prompt-based creation approach we presented in this paper to build environments previously specified for the benchmark, for example some of the scenarios we listed in Section \ref{sec:scenarios}, as well as simpler game-level like environments (e.g. a fantasy tavern full of NPCs, an ancient maze full of traps, etc.). These creations would then have to be compared against those of a professional game dev team. The evaluation criteria would then have to take into account test gamer audience satisfaction, the time spent on the task and the amount of money spent to achieve the performance (comparing the cost of the API calls and the non-expert hourly remuneration vs the resources and salary needed for such a dev team). Such a study is most likely economically unfeasible, but perhaps a cheaper version could be achieved with support of the open source and game modding communities.

If the motivation is changing the way VR is used, so that spontaneous invention becomes part of the experience more often, then something new is being brought about so there is nothing old to serve in comparison. The introduction of prompt-based creation of virtual worlds opens up completely new modes of interaction in VR and its impact cannot be quantitatively evaluated.  The only criteria available is success in either the marketplace or in informal use.

% ------------------------------------------------------------------------

% Perplexity, HumanEval, BLEU, etc can work in a constrained setting, but what about an open sandbox environment for creativity. How should generation of images and 3d content be evaluated? 

% Evolving and partially subjective

% In our case it might be even harder due to the multimodal structure here

% "low end study" and expensive:
% Could run a study to have people using these tools
% - eventually check gamer satisfaction

% community version: might be cheaper

\section{Discussion \& Conclusion}

Models like Codex enable spontaneous content creation within a VR experience and will be crucial components of a compelling metaverse. We have shown in this paper that prompt-based methods are not only useful in speeding up development, but that they can become integral part of gameplay.  We have demonstrated that Codex can be not only applied to speaking complex scenes into existence with our Holodeck demonstration, or spawning tools on user's hands,  but also enables novel gameplay mechanics. In our main example of Codex VR Pong, we have shown that it is possible to utilize LLMs to enable unplanned creative interactions between arbitrary objects co-created with the AI model. We have outlined a vision for what prompt-based AI co-creation in VR can be like and the type of high-impact applications it might have.

We have identified a list of technical challenges while working on our example demonstrations. Together with our thoughts on how to evaluate such AI co-creation, these suggest many directions for future work. Future code generating models are going to unlock even more impressive world building capabilities, especially in terms of building complex visual scenes. It is clear however, that such code generating models have to be combined with text-to-mesh models to achieve the flexibility of speaking \emph{anything} into existence -- the first step towards this would be to combine Codex with a Dreamfusion like model.  For believable VR worlds, future work has to be focused on not only generating static content, but also including animations and physically sensible modes of interaction with the created objects. 

AI co-creation has the potential to transform the way we interact with virtual worlds and bring about the early VR hopes of speaking shared environments into existence, allowing for novel modes of communicating ideas. It remains to be seen how the various technical challenges can be resolved and the cost associated with such models brought down to the level allowing for wide adoption of this technology.

\section*{acknowledgments}
The authors wish to thank Ryan Volum, Jennifer Marsman, Judith Amores, Abby Harrison, Jason Carter, Sebastian Vanderberghe, Ian Norris, Valentine Kozin and Marek Kowalski for the lengthy discussions on speaking the world into existence. 

\bibliographystyle{abbrv-doi}

\bibliography{references}

\appendix
\section{Context for prompts}
Here we collect the context appended to prompts for our experiments. Codex and GPT-3 are highly capable in the few-shot regime, in which we provide a few examples of desired operation.

\subsection{Prompt to GPT-3 for creative collisions}
This is a magical game like ping pong, in which the players can change both the ball and their paddles, when the transformed ball object hits the transformed paddle, it changes the ball according to how you'd expect those two objects to interact.

When a spawned loaf of bread collides with spawned cheese it spawns A sandwich object.

When a spawned pen collides with spawned paper it spawns a notebook object.

When spawned meat collides with a spawned clock it spawns a bacteria object.

When a music note object collides with a cube object it spawns an instrument.

When water object collides with air object it spawns ice. 

When a tree collides with a clock, it spawns a dead tree.

When an egg collides with a clock, it spawns a chicken.

When a cube collides with a wheel, it spawns a car.

When an egg collides with a frying pan, it spawns a fried egg.

When a balloon collides with a pin, it spawns a popped balloon.

When a bread collides with a clock, it spawns a moldy bread.

When a caterpillar collides with a clock, it spawns a butterfly.

When water collides with fire, it spawns steam.

When seed collides with water, it spawns a plant.

When egg collides with clock, it spawns

\subsection{Context for Holodeck}

/* This document contains natural language commands and the Unity $C\#$ code needed to accomplish them .
 Specifically this code modifies and adds objects to the scene to implement a Holodeck from Star Trek.

The starting scene is am empty 10x10x10 room and the objects in the scene are [Floor, Ceiling, North Wall, East Wall, South Wall, West Wall]. The floor is at height y = 0, the ceiling at y = 10. The walls are at (0,0,5), (5,0,0), (0,0,-5) and (-5,0,0). No objects should be placed outside of these bounds.

New models can be added to the scene by using the Sketchfab API (which has to be instantiated with SketchfabLoader loader = gameObject.GetComponent<SketchfabLoader>();) by calling the method SketchfabLoader.LoadModel("model\_name") and can be placed next to an existing model with SketchfabLoader.PlaceModelNextTo(existingModel, newModel, direction), where direction is a Vector3 specifying direction in which to place the new object. SketchfabLoader.MoveModel(model, location) moves the model by updating it's transform to the location specified. The Sketchfab API uses a custom Model class instead of GameObject for asynchronous operations. These models are scaled to fit in a 1x1x1 box, so they need to be rescaled according to real world size using the method SketchfabLoader.ScaleModel(model, size); where size is a float in meters for the largest extension of the object.  Physics can be added to the objects by calling the function SketchfabLoader.AddPhysics(model, mass), where mass is a float number for the mass in kilograms. Previously added objects can be all removed by calling SketchfabLoader.DestroyLoadedObjects(). 

Primitive geometric shapes can be added as usual with CreatePrimitive, e.g. GameObject.CreatePrimitive(PrimitiveType.Cube) to create a cube. Same goes for the usage of any of the standard Unity classes that do not have to download data asynchronously from the cloud. */

/* Example: Add a model of a Computer Desk in front of the north wall and a Flashlight on top of it. Rescale them to the average real world size in meters, and add physics to them with the average mass of such an object.  */

using UnityEngine;

using System.Collections;

public class AddDeskAndFlashlight : MonoBehaviour \{

	void Start () \{
	
		SketchfabLoader loader = gameObject.GetComponent<SketchfabLoader>();
		
		SketchfabLoader.Model northWall = new SketchfabLoader.Model(GameObject.Find("North Wall"));		
		
		SketchfabLoader.Model computerDesk = loader.LoadModel("Computer Desk");
		
		loader.ScaleModel(computerDesk, 1.77f);
		
		loader.AddPhysics(computerDesk, 30); //The average computer desk weighs 30 kg
		
		loader.PlaceModelNextTo(northWall, computerDesk, new Vector3(0, 0, -0.5f));		
		
		SketchfabLoader.Model flashlight = loader.LoadModel("Flashlight");
		
		loader.ScaleModel(flashlight, 0.2f);
		
		loader.PlaceModelNextTo(computerDesk, flashlight, new Vector3(0,1,0));
		
		loader.AddPhysics(flashlight, 0.25f); // The average flashlight weighs 0.25 kg
		
	\}
	
\}

\section{Context for Speaking tools onto hands} 

/* This document contains natural language commands and the Unity $C\#$ code needed to accomplish them */
/* Specifically this code modifies the leap motion hands in the scene by adding models as children to different joints */

/* The left hand in the scene contains the joints [L\_Wrist, L\_Palm, L\_thumb\_meta, L\_thumb\_a, L\_thumb\_b, L\_thumb\_end, L\_index\_meta, L\_index\_b, L\_index\_c, L\_index\_end, L\_middle\_meta, L\_middle\_a, L\_middle\_b, L\_middle\_c, L\_middle\_end, L\_ring\_meta, L\_ring\_a, L\_ring\_b, L\_ring\_c, L\_ring\_end, L\_pinky\_meta, L\_pinky\_a, L\_pinky\_b, L\_pinky\_c, L\_pinky\_end] */

/* The right hand is similar and contains the joints [R\_Wrist, R\_Palm, R\_thumb\_meta, R\_thumb\_a, R\_thumb\_b, R\_thumb\_end, R\_index\_meta, R\_index\_b, R\_index\_c, R\_index\_end, R\_middle\_meta, R\_middle\_a, R\_middle\_b, R\_middle\_c, R\_middle\_end, R\_ring\_meta, R\_ring\_a, R\_ring\_b, R\_ring\_c, R\_ring\_end, R\_pinky\_meta, R\_pinky\_a, R\_pinky\_b, R\_pinky\_c, R\_pinky\_end]

/* Add a Medical Saw model to the right leap motion wrist */

using UnityEngine;

using System.Collections;

public class MedicalSawWrist : MonoBehaviour \{

	void Start () \{
	
		GameObject rightWrist = GameObject.Find("R\_Wrist");	
		
		GameObject medicalSaw = Instantiate(Resources.Load("Medical Saw") as GameObject,rightWrist.transform);
		
		medicalSaw.name = "Medical Saw";
		
	\}
	
\}

\end{document}